\documentclass[12pt,epsf]{article}
\usepackage{graphicx}
\usepackage{epsfig}
\begin{document}

\def\a{\alpha}
\def\b{\beta}
\def\c{\varepsilon}
\def\d{\delta}
\def\e{\epsilon}
\def\f{\phi}
\def\g{\gamma}
\def\h{\theta}
\def\k{\kappa}
\def\l{\lambda}
\def\m{\mu}
\def\n{\nu}
\def\p{\psi}
\def\q{\partial}
\def\r{\rho}
\def\s{\sigma}
\def\t{\tau}
\def\u{\upsilon}
\def\v{\varphi}
\def\w{\omega}
\def\x{\xi}
\def\y{\eta}
\def\z{\zeta}
\def\D{\Delta}
\def\G{\Gamma}
\def\H{\Theta}
\def\L{\Lambda}
\def\F{\Phi}
\def\P{\Psi}
\def\S{\Sigma}
\def\Vec#1{\mbox{\boldmath $#1$}}
\def\pT#1{\Vec{p}_{\rm T}^{#1}}

\def\o{\over}
\def\beq{\begin{eqnarray}}
\def\eeq{\end{eqnarray}}
\newcommand{\lsim}{\raisebox{0.6mm}{$\, <$} \hspace{-3.0mm}\raisebox{-1.5mm}{\em $\sim \,$}}
\newcommand{\gsim}{\raisebox{0.6mm}{$\, >$} \hspace{-3.0mm}\raisebox{-1.5mm}{\em $\sim \,$}}

\newcommand{\vev}[1]{ \left\langle {#1} \right\rangle }
\newcommand{\bra}[1]{ \langle {#1} | }
\newcommand{\ket}[1]{ | {#1} \rangle }
\newcommand{\EV}{ {\rm eV} }
\newcommand{\KEV}{ {\rm keV} }
\newcommand{\MEV}{ {\rm MeV} }
\newcommand{\GEV}{ {\rm GeV} }
\newcommand{\TEV}{ {\rm TeV} }
\def\diag{\mathop{\rm diag}\nolimits}
\def\Spin{\mathop{\rm Spin}}
\def\SO{\mathop{\rm SO}}
\def\O{\mathop{\rm O}}
\def\SU{\mathop{\rm SU}}
\def\U{\mathop{\rm U}}
\def\Sp{\mathop{\rm Sp}}
\def\SL{\mathop{\rm SL}}
\def\tr{\mathop{\rm tr}}

\def\IJMP{Int.~J.~Mod.~Phys. }
\def\MPL{Mod.~Phys.~Lett. }
\def\NP{Nucl.~Phys. }
\def\PL{Phys.~Lett. }
\def\PR{Phys.~Rev. }
\def\PRL{Phys.~Rev.~Lett. }
\def\PTP{Prog.~Theor.~Phys. }
\def\ZP{Z.~Phys. }


\baselineskip 0.7cm

\begin{titlepage}

\begin{flushright}
UT-08-17
\\
IPMU-08-0029
\end{flushright}

\vskip 1.35cm
\begin{center}
{\large \bf
A Measurement of Neutralino Mass at the LHC\\
in Light Gravitino Scenarios
}
\vskip 1.2cm
Koichi Hamaguchi$^{1,2}$, Eita Nakamura$^1$ and Satoshi Shirai$^1$
\vskip 0.4cm

{\it $^1$  Department of Physics, University of Tokyo,\\
     Tokyo 113-0033, Japan\\
$^2$ Institute for the Physics and Mathematics of the Universe, 
University of Tokyo,\\ Chiba 277-8568, Japan}

\vskip 1.5cm

\abstract{
We consider supersymmetric (SUSY) models in which a very light gravitino is the lightest SUSY particle.
Assuming that a neutralino is the next-to-lightest SUSY particle,
we present a measurement of the neutralino mass at the LHC in two photons + missing energy events,
which is based on the $M_{\rm T2}$ method.
It is a direct measurement of the neutralino mass itself, independent of other SUSY particle masses
and patterns of cascade decays before the neutralino is produced.
}
\end{center}
\end{titlepage}

\setcounter{page}{2}
Among various supersymmetric (SUSY) models, those 
with an ultralight gravitino of mass $m_{3/2} \lsim {\mathcal O}(10)$ eV 
are very attractive, since they are completely free from notorious
gravitino problems~\cite{grav10eV}.
In this letter, we assume a neutralino is the next-to-lightest SUSY particle (NLSP), 
and present a measurement of its mass at the LHC. 
It is based on the so-called  $M_{\rm T2}$ method~\cite{Lester-Summers}.
We show that this method can directly determine the neutralino mass, 
independently of other SUSY particle masses, 
and it does not rely on specific patterns of cascade decays before the neutralino is produced.

In the scenario considered here, essentially all the SUSY events will end up with two 
neutralino NLSPs,\footnote{We assume $R$-parity conservation.} each of which 
then dominantly decays into a gravitino and a photon.\footnote{We do not discuss the case in which
the neutralino mainly decays into a Higgs/$Z$-boson and a gravitino.}
We assume that the decay length of the NLSP neutralino is so short that  
the decay occurs inside the detector and the photons' momenta are measured well.
Therefore, the main signature at the LHC will be two high transverse momentum photons 
and a large missing transverse momentum.
If such a signal will indeed be discovered, 
one of the most natural candidates for the underlying model
is a SUSY model with a gravitino LSP and a neutralino NLSP.

Furthermore,
from the prompt decay of the neutralino, we can assume that the gravitino is very light,
essentially massless for the following discussion. 
This is because the NLSP decay length
is proportional to the gravitino mass squared as
\begin{equation}
c \tau_{\rm NLSP}\sim 20 \mu{\rm m}
\left(\frac{m_{3/2}}{1~{\rm eV}}\right)^2
\left(\frac{m_{\rm NLSP}}{100~{\rm GeV}}\right)^{-5},
\end{equation}
and a heavier gravitino ($m_{3/2}>{\cal O}(1)$ keV)
would make the neutralino decay outside the detector.\footnote{For a moderate gravitino mass
corresponding $c\tau_{\rm NLSP}={\cal O}(10)~{\rm cm}-{\cal O}(10)~{\rm m}$, the neutralino decay causes
``non-pointing" photons~\cite{Kawagoe:2003jv}. The present method of the neutralino mass
determination may also work in this case.}
This indirect information of the massless LSP plays
a crucial role in the NLSP mass determination.

Let us start by briefly explaining the $M_{\rm T2}$ method~\cite{Lester-Summers}. 
Suppose that there is a particle $A$ which promptly decays by the process $A\to B+X$, 
where $B$ is a visible (standard-model) particle and $X$ is a neutral and undetected particle.
When two $A\,$s are produced in a collider, we can measure the two $B$s' transverse momenta
 $\pT{B,1}$, $\pT{B,2}$ 
and the missing transverse momentum
$\pT{\rm miss}=\pT{X,1}+\pT{X,2}$.\footnote{We 
assume that the missing $\Vec{p}_{\rm T}$ is dominantly caused by the two $X$s 
and the contribution of other sources of missing $\Vec{p}_{\rm T}$ are negligible.}
The $M_{\rm T2}$ variable is then given by
\beq
\label{eq:mt2}
(M_{\rm T2})^2
\equiv\mathop{\rm min}_{\pT{{\rm miss},1}+\pT{{\rm miss},2}=
\pT{\rm miss}}\Big[{\rm max}\Big\{\big(M_{\rm T}^{(1)}\big)^2,
\big(M_{\rm T}^{(2)}\big)^2\Big\}\Big],
\eeq
where the minimization is taken over all possible momentum splittings, and
\beq
\label{eq:mt}
(M_{\rm T}^{(i)})^2
=
m_B^2 + m_X^2 + 2(E_{\rm T}^{{\rm miss},i} E_{\rm T}^{B,i}
-
\pT{{\rm miss},i}\cdot\pT{B,i})
\quad{\rm for}\quad i=1,2
\eeq
with $E_{\rm T}^{{\rm miss},i}\equiv\sqrt{m_X^2+|\pT{{\rm miss},i}|^2}$
and $E_{\rm T}^{B,i}\equiv\sqrt{m_B^2+|\pT{B,i}|^2}$.
This $M_{\rm T2}$ variable is designed to have the endpoint at $m_A$ 
when we input the correct value of $m_X$.
However, in general, the mass $m_X$ of the missing particle $X$ is unknown, and therefore
one can obtain only a relation between $m_X$ and $m_A$.\footnote{See also recent developments in the $M_{\rm T2}$ method~\cite{recentMT2}.}

A crucial point in the scenario considered here 
($\{A, B, X\} =$ $\{$neutralino, $\gamma$, gravitino$\}$) is that 
we can assume the massless LSP, $m_X=m_{3/2}=0$, as discussed above.
Therefore, we can directly determine the NLSP mass by the $M_{\rm T2}$ method.
As we show in the Appendix, the $M_{\rm T2}$ variable in this case is analytically expressed as~\footnote{
For completeness,
we also show an analytic expression of $M_{\rm T2}$ 
for the case of massive LSP ($m_X\ne 0$) in the Appendix.}
\begin{eqnarray}
\big(M_{\rm T2}\big)^2 &=& 
\left\{
\begin{array}{lll}
2 p_1 p_2 z & {\rm for} & c_1<0 ~{\rm or}~ c_2 <0
\\
0 & {\rm for} & c_1\ge 0 ~{\rm and}~ c_2\ge 0
\end{array}
\right.
,
\label{eq:mt2nlsp}
\end{eqnarray}
where $p_1\equiv|\pT{\gamma,1}|$, $p_2\equiv|\pT{\gamma,2}|$, 
$c_1$ and $c_2$ are given by 
\begin{eqnarray}
\pT{\rm miss} &=& c_1\pT{\gamma,1}+c_2\pT{\gamma,2},
\end{eqnarray}
and
$z$ is a real positive solution of the following equations:
\begin{eqnarray}
4(a-b)^2 &=& \big(\sqrt{2(a+b)+3}-1\big)\big(\sqrt{2(a+b)+3}+3\big)^3\,,
\nonumber
\\
a &=& \frac{1}{r^{\frac{1}{3}}}\bigg[\frac{r}{2}-\cos\theta
+c_1 \sin^2\theta
\frac{1}{z}\bigg],
\nonumber
\\
b &=& r^{\frac{1}{3}}\bigg[\frac{1}{2r}-\cos\theta
+c_2 \sin^2\theta
\frac{1}{z}\bigg],
\nonumber
\\
r &=&\frac{p_2}{p_1},
\quad
\cos\theta = \frac{\pT{\gamma,1}\cdot\pT{\gamma,2}}{p_1 p_2}.
\end{eqnarray}
Note that $M_{\rm T2}$ is completely defined by the missing transverse momentum and 
photon momenta, independently of other kinematical variables.
We should also emphasize that the present method does not rely on a direct pair-production of the NLSPs, 
i.e., we do not assume back-to-back transverse momenta of the NLSPs,
$\pT{\rm miss}+\pT{\gamma,1}+\pT{\gamma,2}=0$.
In the following, we show how this method works at the LHC, by taking explicit examples of 
gauge mediated SUSY breaking (GMSB) models 
which realize the mass spectrum with an ultralight gravitino LSP and a neutralino NLSP.


We consider two gauge mediation models for a demonstration.
In the following, mass spectrums are calculated by ISAJET 7.72~ \cite{ISAJET} and
we use programs Herwig 6.5~\cite{HERWIG6510} and AcerDET-1.0~\cite{RichterWas:2002ch}
 to simulate LHC signatures.

\begin{figure}[t]
\begin{center}
\scalebox{1}{
\begin{picture}(220,200.12)(0,0)
\put(5,5.9945){\line(1,0){20}}
\put(10,8.9945){$\scriptstyle h^0$}
\put(5,105.87){\line(1,0){20}}
\put(5,105.465){\line(1,0){20}}
\put(5,105.941){\line(1,0){20}}
 \put(5,108.941){$\scriptstyle H^{\pm}$}
 \put(5,98.8695){$\scriptstyle A^0,H^0$}
\put(37,17.793){\line(1,0){20}}
\put(42.,21.793){$\scriptstyle {\tilde{\chi}^0_1}$}
\put(37,44.9685){\line(1,0){20}}
\put(37.,36.9685){$\scriptstyle {\tilde{\chi}^0_2,\tilde{\chi}^0_3}$}
\put(37,46.2375){\line(1,0){20}}
\put(37,53.1705){\line(1,0){20}}
\put(42.,57.1705){$\scriptstyle {\tilde{\chi}^0_4}$}
\put(69,45.7355){\line(1,0){20}}
\put(74.,36.7355){$\scriptstyle {\tilde{\chi}^{\pm}}_1$}
\put(69,52.313){\line(1,0){20}}
\put(74.,56.313){$\scriptstyle {\tilde{\chi}^{\pm}_2}$}
\put(101,30.63){\line(1,0){20}}
\put(106,34.63){$\scriptstyle {\tilde g}$}
\put(133,207.863){\line(1,0){20}}
\put(133,207.824){\line(1,0){20}}
\put(133,211.863){$\scriptstyle \tilde d_L\,\tilde u_L$}
\put(133,185.325){\line(1,0){20}}
\put(133,186.641){\line(1,0){20}}
\put(133,177.325){$\scriptstyle \tilde d_R\,\tilde u_R$}
\put(133,95.98){\line(1,0){20}}
\put(133,95.584){\line(1,0){20}}
\put(138,87.584){$\scriptstyle \tilde \nu_e$}
\put(138,99.98){$\scriptstyle \tilde e_L$}
\put(133,45.487){\line(1,0){20}}
\put(138,50.0){$\scriptstyle \tilde e_R $}
\put(165,163.232){\line(1,0){20}}
\put(170,155.232){$\scriptstyle \tilde t_1$}
\put(165,200.12){\line(1,0){20}}
\put(165,184.816){\line(1,0){20}}
\put(170,188.816){$\scriptstyle \tilde b_1$}
\put(165,198.896){\line(1,0){20}}
\put(165,202.896){$\scriptstyle \tilde b_2\,\tilde t_2$}
\put(165,45.9365){\line(1,0){20}}
\put(170,49.9365){$\scriptstyle \tilde \tau_1$}
\put(165,95.7455){\line(1,0){20}}
\put(170,99.7455){$\scriptstyle \tilde \tau_2$}
\put(165,95.4025){\line(1,0){20}}
\put(170,87.4025){$\scriptstyle \tilde \nu_{\tau}$}
\put(260,100.06){\small Mass }
\put(260,90.06){$\scriptstyle {\rm (GeV)}$ }
\put(220,-10){\vector(0,1){225.12}}
\put(220,0){\line(1,0){10}}
\put(232,-2){$\scriptstyle 0$}
\put(220,25){\line(1,0){10}}
\put(232,23){$\scriptstyle 500$}
\put(220,50){\line(1,0){10}}
\put(232,48){$\scriptstyle 1000$}
\put(220,75){\line(1,0){10}}
\put(232,73){$\scriptstyle 1500$}
\put(220,100){\line(1,0){10}}
\put(232,98){$\scriptstyle 2000$}
\put(220,125){\line(1,0){10}}
\put(232,123){$\scriptstyle 2500$}
\put(220,150){\line(1,0){10}}
\put(232,148){$\scriptstyle 3000$}
\put(220,175){\line(1,0){10}}
\put(232,173){$\scriptstyle 3500$}
\put(220,200){\line(1,0){10}}
\put(232,198){$\scriptstyle 4000$}
\put(220,0){\line(1,0){5}}
\put(220,12.5){\line(1,0){5}}
\put(220,25){\line(1,0){5}}
\put(220,37.5){\line(1,0){5}}
\put(220,50){\line(1,0){5}}
\put(220,62.5){\line(1,0){5}}
\put(220,75){\line(1,0){5}}
\put(220,87.5){\line(1,0){5}}
\put(220,100){\line(1,0){5}}
\put(220,112.5){\line(1,0){5}}
\put(220,125){\line(1,0){5}}
\put(220,137.5){\line(1,0){5}}
\put(220,150){\line(1,0){5}}
\put(220,162.5){\line(1,0){5}}
\put(220,175){\line(1,0){5}}
\put(220,187.5){\line(1,0){5}}
\end{picture} }
\end{center}
\caption{Mass spectrum of SIGM}
\label{fig:spectrumSIGM}
\end{figure}
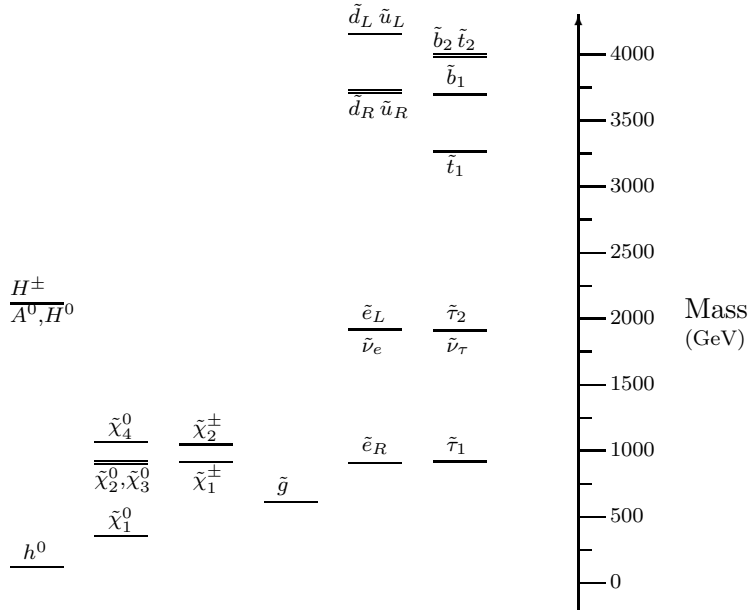

The first example is a strongly interacting gauge mediation (SIGM) model ~\cite{Hamaguchi:2008yu}, in which
the NLSP is a neutralino.
We take the same SIGM parameters as the example in Sec. 4 of Ref.~\cite{Hamaguchi:2008yu}.
The mass spectrum is shown in Fig.~\ref{fig:spectrumSIGM}.
The masses of the lightest neutralino and gravitino are 356 GeV and 10 eV, respectively.

We take the events cuts as follows:
\begin{itemize}
\item $\ge 4$ jets with $p_{\rm T}>50$ GeV and $p_{\rm T,1,2}>100$ GeV.
\item $\ge 2$ photons with $p_{\rm T}>20$ GeV.
\item $M_{\rm eff}>500$ GeV, where
\begin{equation}
M_{\rm eff} = \sum_{\rm jets}^{4} p_{{\rm T}j} + p_{\rm T}^{\rm miss}.
\end{equation} 
\item $p_{\rm T}^{\rm miss}>0.2 M_{\rm eff}$.
\end{itemize}
Under these cuts, we see that the standard-model backgrounds are almost negligible.

\begin{figure}[t]
\begin{tabular}{cc}
\begin{minipage}{0.55\hsize}
\begin{center}
\epsfig{file=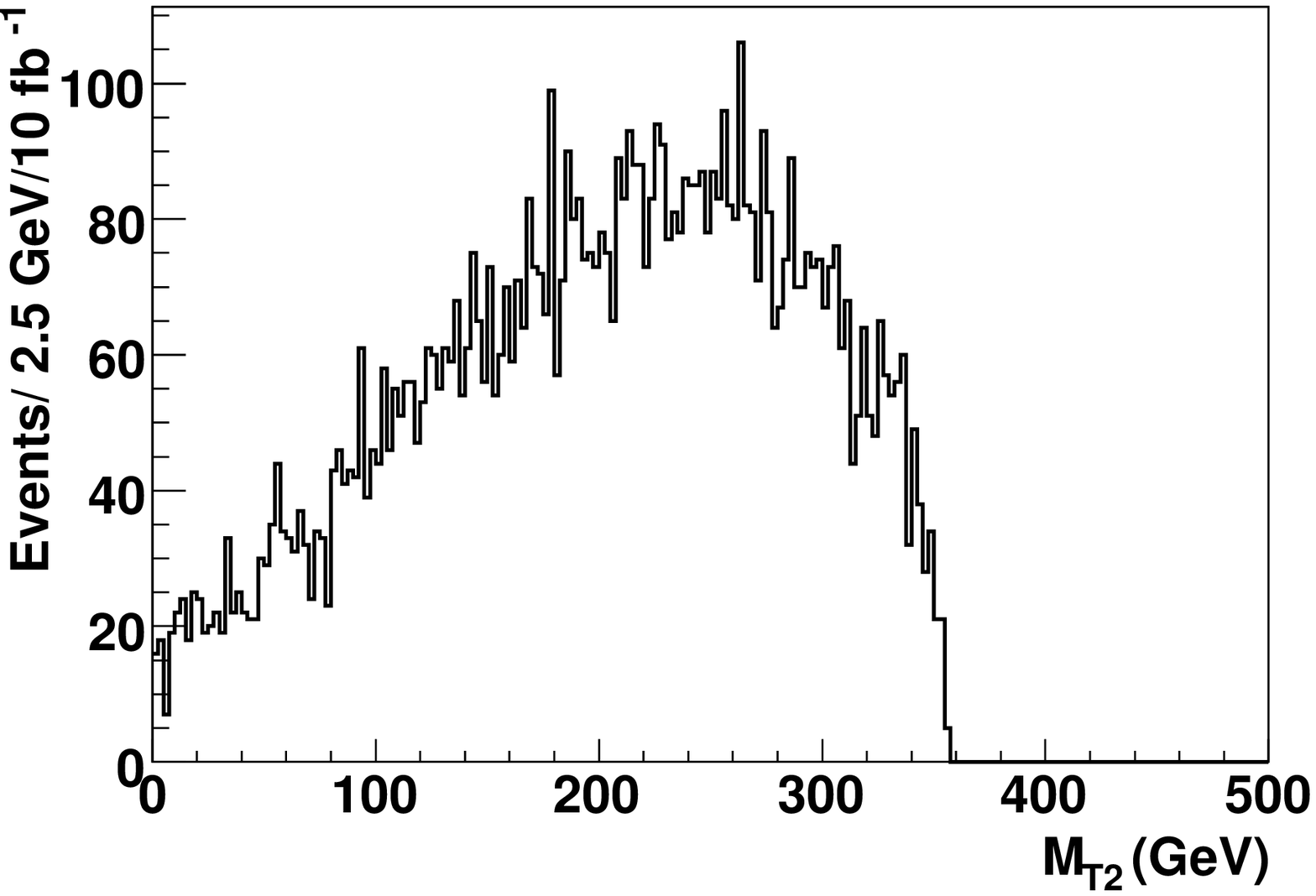 ,scale=.45,clip}
(a)
\end{center}
\end{minipage}
\begin{minipage}{0.55\hsize}
\begin{center}
\epsfig{file=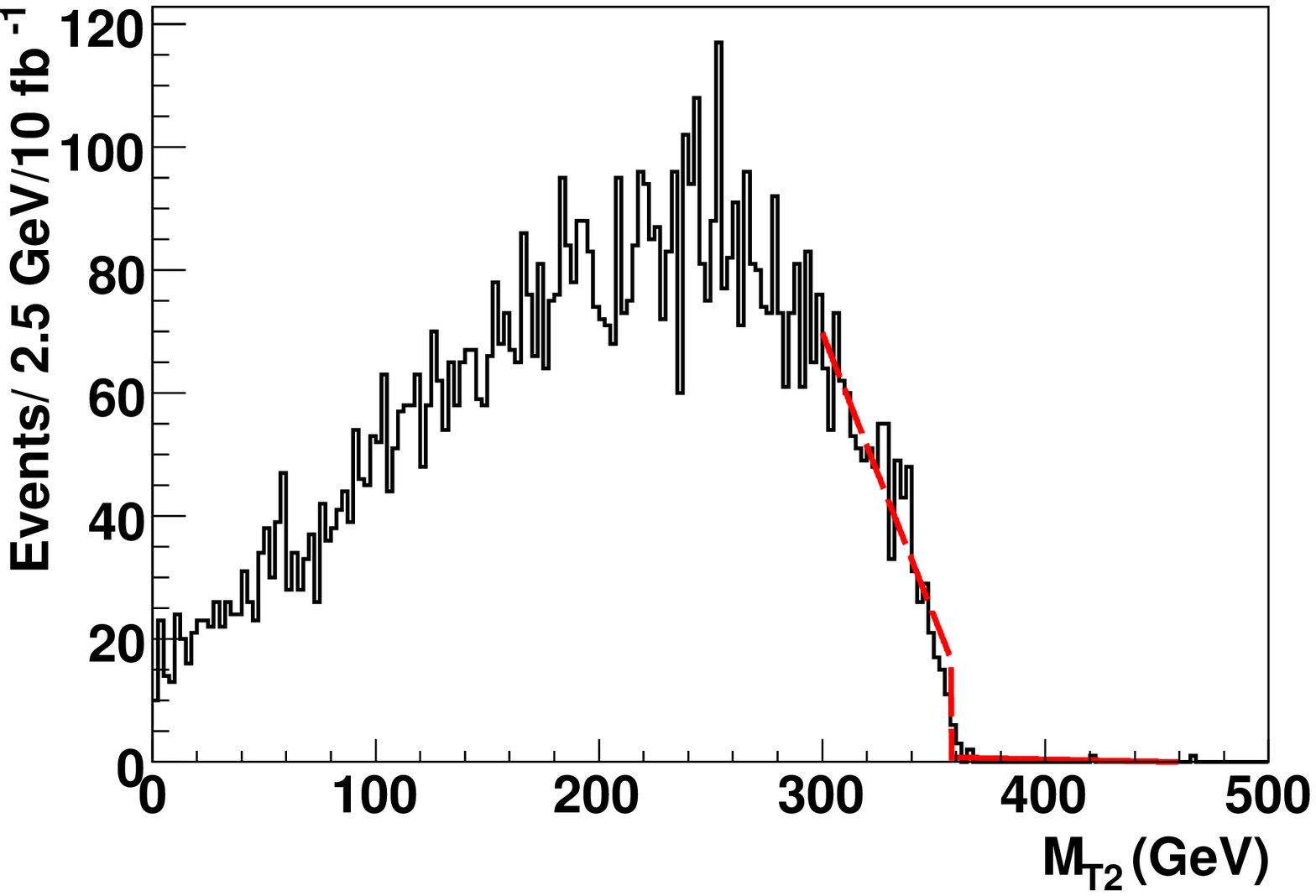 ,scale=.45,clip}
(b)
\end{center}
\end{minipage}
\end{tabular}
\caption{A distribution of $M_{\rm T2}$ for the SIGM example. (a): parton level signature. 
(b): detector level signature.
}
\label{fig:SIGM}
\end{figure}

In Fig.~\ref{fig:SIGM}-(a), a parton level distribution of $M_{\rm T2}$ is shown for an integrated luminosity of 10 fb$^{-1}$.
Here, we take the sum of gravitino and neutrino transverse momenta 
as the parton level missing $\Vec{p}_{\rm T}$.
As discussed in Ref.~\cite{Hamaguchi:2008yu}, very little number of leptons are produced in the SIGM.
Therefore, missing $\Vec{p}_{\rm T}$ is due to almost only gravitinos and the assumption that $\pT{\rm miss}
= \pT{\rm LSP1}+\pT{\rm LSP2}$ is satisfied. 
There is a clear edge at $M_{\rm T2} \simeq m_{\tilde{\chi}^0_1}=356$ GeV.

In Fig.~\ref{fig:SIGM}-(b), we show a distribution of $M_{\rm T2}$ after taking account of detector effects.
In order to extract the point of the edge, we use a simple fitting function;
\begin{equation}
f(x) = (ax+b)\theta(-x+M) + (cx+d)\theta(x-M), \label{eq:fit}
\end{equation}
where $\theta(x)$ is the step function and $a, b, c, d$ and $M$ are fitting parameters. 
We fit the data with $f(x)$ over $300 \le M_{\rm T2} \le 500$ GeV and find
\begin{equation}
m_{\tilde{\chi}^0_1} = 357 \pm 3~ {\rm GeV}.
\end{equation}
Here, the estimation of the error is done by `eye' because of lack of information 
on the shape of the $M_{\rm T2}$ distribution. 
The estimation that $m_{\tilde{\chi}^0_1} = 357 \pm 3$ GeV is very good agreement with the true value  $m_{\tilde{\chi}^0_1} = 356$ GeV.

\begin{figure}[t]
\begin{center}
\scalebox{1}{
\begin{picture}(220,217.764)(0,0)
\put(5,22.586){\line(1,0){20}}
\put(10,25.586){$\scriptstyle h^0$}
\put(5,106.818){\line(1,0){20}}
\put(5,106.366){\line(1,0){20}}
\put(5,107.982){\line(1,0){20}}
 \put(5,110.982){$\scriptstyle H^{\pm}$}
 \put(5,99.818){$\scriptstyle A^0,H^0$}
\put(37,27.864){\line(1,0){20}}
\put(42.,31.864){$\scriptstyle {\tilde{\chi}^0_1}$}
\put(37,52.618){\line(1,0){20}}
\put(42.,56.618){$\scriptstyle {\tilde{\chi}^0_2}$}
\put(37,84.552){\line(1,0){20}}
\put(42.,76.552){$\scriptstyle {\tilde{\chi}^0_3}$}
\put(37,88.626){\line(1,0){20}}
\put(42.,92.626){$\scriptstyle {\tilde{\chi}^0_4}$}
\put(37,88.626){\line(1,0){20}}
\put(42.,92.626){$\scriptstyle {\tilde{\chi}^0_4}$}
\put(69,52.708){\line(1,0){20}}
\put(74.,56.708){$\scriptstyle {\tilde{\chi}^{\pm}}_1$}
\put(69,88.588){\line(1,0){20}}
\put(74.,92.588){$\scriptstyle {\tilde{\chi}^{\pm}_2}$}
\put(101,167.536){\line(1,0){20}}
\put(106,171.536){$\scriptstyle {\tilde g}$}
\put(133,223.22){\line(1,0){20}}
\put(133,222.64){\line(1,0){20}}
\put(133,227.22){$\scriptstyle \tilde d_L\,\tilde u_L$}
\put(133,212.622){\line(1,0){20}}
\put(133,213.372){\line(1,0){20}}
\put(133,204.622){$\scriptstyle \tilde d_R\,\tilde u_R$}
\put(133,71.83){\line(1,0){20}}
\put(133,69.538){\line(1,0){20}}
\put(138,61.538){$\scriptstyle \tilde \nu_e$}
\put(138,75.83){$\scriptstyle \tilde e_L$}
\put(133,35.01){\line(1,0){20}}
\put(138,39.01){$\scriptstyle \tilde e_R $}
\put(165,194.14){\line(1,0){20}}
\put(170,186.14){$\scriptstyle \tilde t_1$}
\put(165,217.764){\line(1,0){20}}
\put(165,210.94){\line(1,0){20}}
\put(170,202.94){$\scriptstyle \tilde b_1$}
\put(165,214.866){\line(1,0){20}}
\put(165,219.866){$\scriptstyle \tilde b_2\,\tilde t_2$}
\put(165,34.696){\line(1,0){20}}
\put(170,38.696){$\scriptstyle \tilde \tau_1$}
\put(165,71.706){\line(1,0){20}}
\put(170,75.706){$\scriptstyle \tilde \tau_2$}
\put(165,69.134){\line(1,0){20}}
\put(170,61.134){$\scriptstyle \tilde \nu_{\tau}$}
\put(240,108.882){\small Mass }
\put(240,98.882){$\scriptstyle {\rm (GeV)}$ }
\put(220,-10){\vector(0,1){227.764}}
\put(220,0){\line(1,0){10}}
\put(232,-2){$\scriptstyle 0$}
\put(220,40){\line(1,0){10}}
\put(232,38){$\scriptstyle 200$}
\put(220,80){\line(1,0){10}}
\put(232,78){$\scriptstyle 400$}
\put(220,120){\line(1,0){10}}
\put(232,118){$\scriptstyle 600$}
\put(220,160){\line(1,0){10}}
\put(232,158){$\scriptstyle 800$}
\put(220,200){\line(1,0){10}}
\put(232,198){$\scriptstyle 1000$}
\put(220,0){\line(1,0){5}}
\put(220,20){\line(1,0){5}}
\put(220,40){\line(1,0){5}}
\put(220,60){\line(1,0){5}}
\put(220,80){\line(1,0){5}}
\put(220,100){\line(1,0){5}}
\put(220,120){\line(1,0){5}}
\put(220,140){\line(1,0){5}}
\put(220,160){\line(1,0){5}}
\put(220,180){\line(1,0){5}}
\end{picture} }
\end{center}
\caption{Mass spectrum of SPS8}
\label{fig:spectrumSPS}
\end{figure}
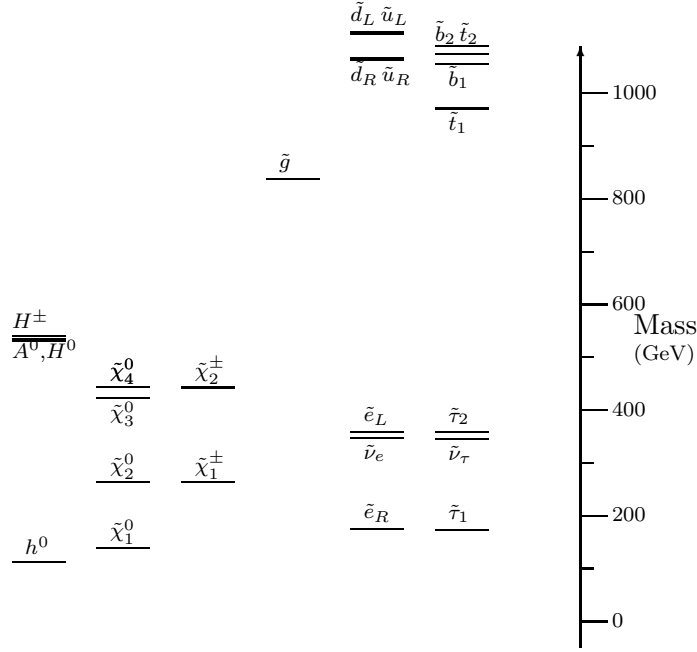

Next we show another example. 
We study the Snowmass benchmark point SPS8~\cite{Allanach:2002nj}, 
which is a minimal gauge mediation model with a neutralino NLSP.
In Fig.~\ref{fig:spectrumSPS}, SPS8 mass spectrum is shown.
The masses of the lightest neutralino and gravitino are 139 GeV and 4.8 eV, respectively.

\begin{figure}[thbp]
\begin{tabular}{cc}
\begin{minipage}{0.55\hsize}
\begin{center}
\epsfig{file=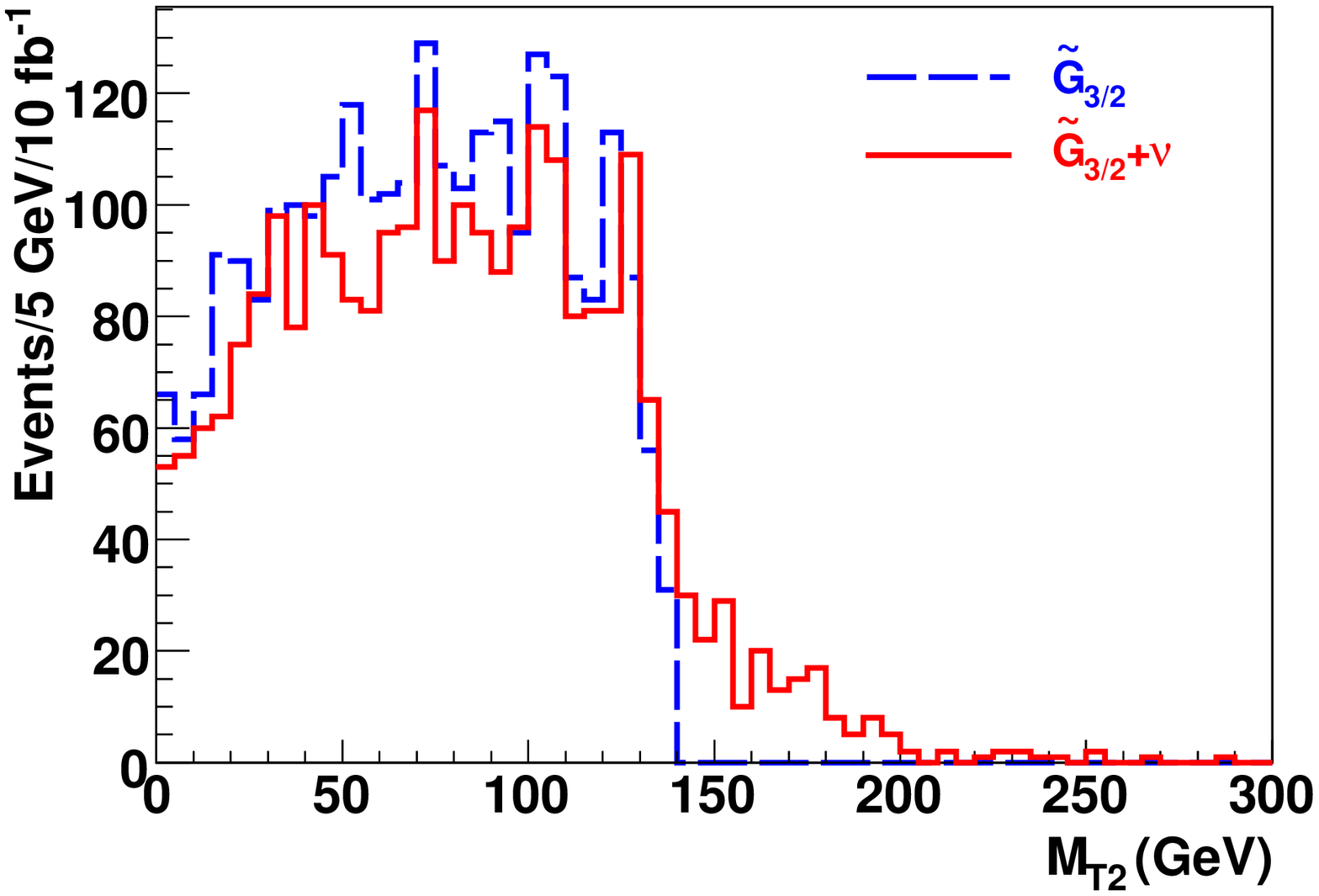 ,scale=.45,clip}
(a)
\end{center}
\end{minipage}
\begin{minipage}{0.55\hsize}
\begin{center}
\epsfig{file=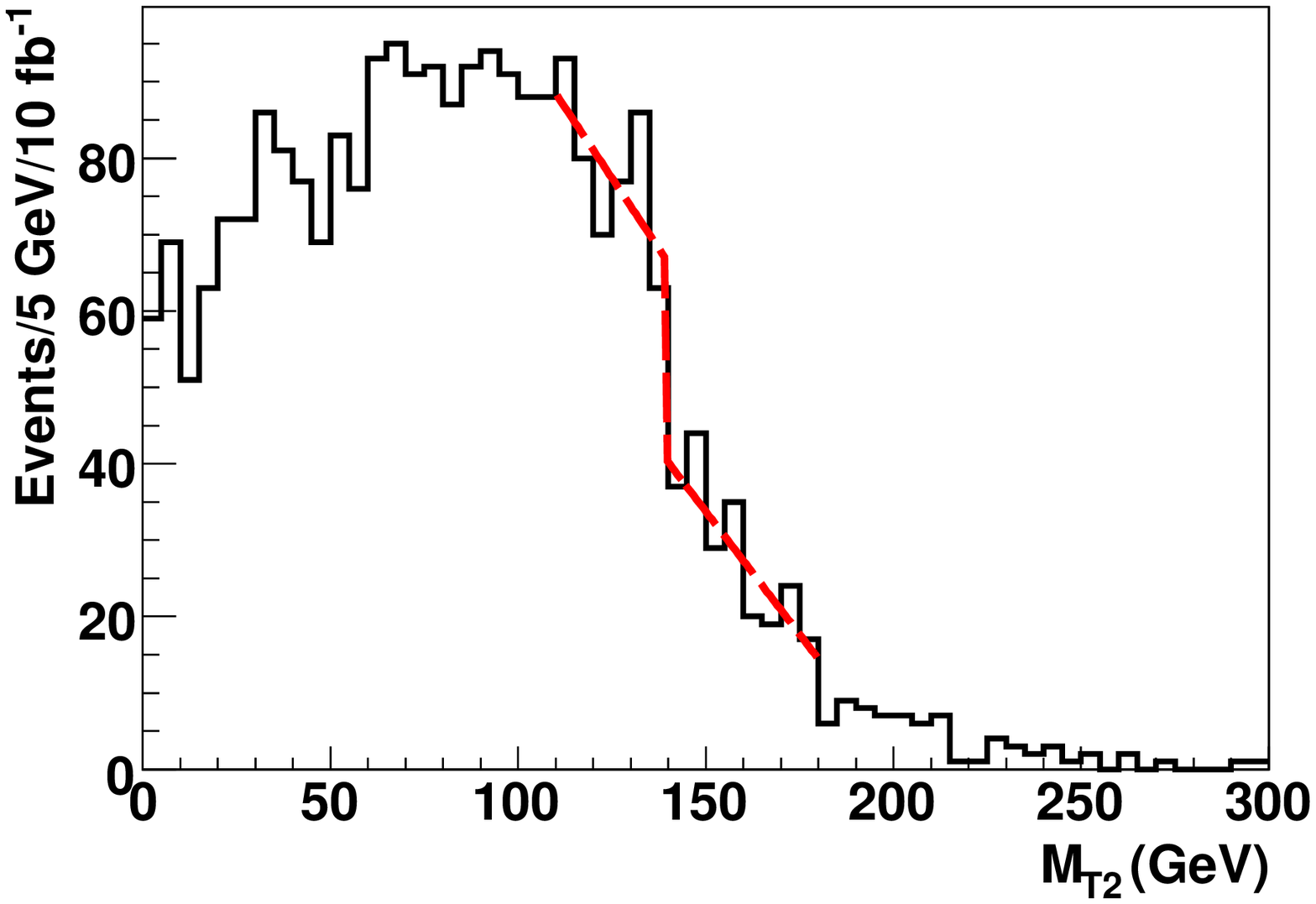 ,scale=.45,clip}
(b)
\end{center}
\end{minipage}
\end{tabular}
\caption{A distribution of $M_{\rm T2}$ for the SPS8. (a): parton level signature. 
The blue and dashed line represents the case that $\pT{\rm miss} = \sum_{\rm gravitino}\Vec{p}_{\rm T}$ and
the red and solid line  $\pT{\rm miss} = \sum_{\rm gravitino}\Vec{p}_{\rm T}+\sum_{\rm neutrino}\Vec{p}_{\rm T}$.
(b): detector level signature.
}
\label{fig:SPS}
\end{figure}
In Fig.~\ref{fig:SPS}-(a), a parton level distribution of $M_{\rm T2}$ is shown for an integrated luminosity of 10 fb$^{-1}$.
The event cuts are the same as in the previous SIGM case.
The blue and dashed line represents the case that $\pT{\rm miss} = \sum_{\rm gravitino}\Vec{p}_{\rm T}$ and
the red and solid line  $\pT{\rm miss} = \sum_{\rm gravitino}\Vec{p}_{\rm T}+\sum_{\rm neutrino}\Vec{p}_{\rm T}$.
In SPS8, there are many neutrino production sources.
Hence, we cannot see a clear edge as in the SIGM case. 
However, there is a cliff at $M_{\rm T2} \simeq m_{\tilde{\chi}^0_1} = 139$ GeV.

In Fig.~\ref{fig:SPS}-(b), detector level distribution of $M_{\rm T2}$ is shown.
To get the value of $m_{\tilde{\chi}^0_1}$, we fit the data with $f(x)$ in Eq.~(\ref{eq:fit}) over 
$110\le M_{\rm T2}\le 180$ GeV. Then we get
\begin{equation}
m_{\tilde{\chi}^0_1} = 139 \pm 3~{\rm GeV}.
\end{equation}
The error estimation is done by `eye'.
This value 
agrees with the true value ($m_{\tilde{\chi}^0_1} = 139 ~{\rm GeV}$).

In summary, we have presented a determination of the neutralino mass
for the SUSY models with an ultralight gravitino LSP and a neutralino NLSP,
which may work in the early stage of the LHC.

Though we have considered GMSB models with a neutralino NLSP, our method is applicable to any model
in which the signal events will lead to a pair of cascade decays that result in
\begin{equation}
 \cdots \to {\rm any~cascade~decay}\to A \to B + X,  \label{eq:decay}
\end{equation}
where $B$ is a visible (standard-model) particle and $X$ is a missing particle that is almost massless.
The mass of $A$ is then determined by the two $B$s' momenta and the missing
transverse momentum.

For example, let us consider GMSB models with a slepton NLSP. 
In this case, the slepton, lepton and gravitino correspond to $A$, $B$ and $X$ in Eq.~(\ref{eq:decay}), respectively. 
In addition to leptons from the sleptons' decays, many other leptons are produced in this scenario. 
However, we may see which of observed leptons is produced through the slepton decay by measuring lepton's momentum,
or by detecting a kink of its track for a long-lived slepton. 
In such a case, we can measure the slepton mass with the $M_{\rm T2}$ method as discussed above.
Furthermore, the present method may work in an axino LSP scenario.

\section*{Acknowledgements}

We thank Tsutomu Yanagida for useful discussion.
This work was supported by World Premier International Center Initiative
(WPI Program), MEXT, Japan.
The work by KH is supported by JSPS (18840012).
The work of SS is supported in part by JSPS
Research Fellowships for Young Scientists.

\section*{Appendix}
In this appendix we derive Eq.~(\ref{eq:mt2nlsp}). 
We start from Eqs.~(\ref{eq:mt2}) and (\ref{eq:mt}).
We assume that $B$ is massless.

\paragraph{(i) $m_X=0$ case:}
First, we consider the case that $X$ is a massless particle.
The $M_{\rm T2}$ variable is defined by Eqs.~(\ref{eq:mt2}) and (\ref{eq:mt}) with $m_{B}=m_{X}=0$.
If a momentum splitting is the correct one, i.e.,
$\pT{{\rm miss},1}=\pT{X,1}$ and $\pT{{\rm miss},2}=\pT{X,2}$,
then each transverse mass is smaller than the mass of $A$, $m_A$:
\beq
m_A^2=(p^{B,i}+p^{X,i})^2=2\big(|\pT{X,i}||\pT{B,i}|\cosh(\Delta y^i)-\pT{X,i}\cdot
\pT{B,i}\big)\geq\big(M_{\rm T}^{(i)}\big)^2
\eeq
for $i=1,2$,
where $\Delta y^i$ is the rapidity difference of $B$ and $X$ in each decay chain.
From this, it is clear that
\beq\label{eq:mT2_bound}
M_{\rm T2}\leq m_A.
\eeq

We do not assume the relation
\beq\label{eq:BtoB}
\pT{B,1}+\pT{B,2}=-\pT{\rm miss},
\eeq
which holds in the case of a ``back-to-back'' pair production of $A\,$s.
We may assume that $\pT{B,1}$ and $\pT{B,2}$ are linearly independent and
$\pT{\rm miss}$ can be expressed as
\beq
\pT{\rm miss}=c_1\pT{B,1}+c_2\pT{B,2}.
\eeq
Here, $c_1$ and $c_2$ are real coefficients and they are given by
\beq
c_1=\frac{1}{\sin^2\theta}\bigg[\frac{\pT{\rm miss}\cdot\pT{B,1}}{(p_1)^2}
-\frac{\pT{\rm miss}\cdot\pT{B,2}}{p_1p_2}\cos\theta\bigg],
\eeq
\beq
c_2=\frac{1}{\sin^2\theta}\bigg[\frac{\pT{\rm miss}\cdot\pT{B,2}}{(p_2)^2}-
\frac{\pT{\rm miss}\cdot\pT{B,1}}{p_1p_2}\cos\theta\bigg],
\eeq
where
\beq
p_1\equiv|\pT{B,1}|,\quad p_2\equiv|\pT{B,2}|,\quad
\cos\theta\equiv\frac{\pT{B,1}\cdot\pT{B,2}}{p_1p_2}.
\eeq
The momentum splitting $\pT{{\rm miss},1}$ and $\pT{{\rm miss},2}$
can also be expressed as
\beq
\pT{{\rm miss},1}=(c_1-x)\pT{B,1}+y\pT{B,2},
\eeq
\beq
\pT{{\rm miss},2}=x\pT{B,1}+(c_2-y)\pT{B,2},
\eeq
where $x$ and $y$ are real variables. We rewrite Eq.~(\ref{eq:mt2}) as
\beq
(M_{\rm T2})^2=2p_1p_2\mathop{\rm min}_{x,y\in{\bf R}}
\Big[{\rm max}\Big\{z_1(x,y),z_2(x,y)\Big\}\Big],
\eeq
where
\beq\label{eq:z1}
z_1(x,y)\equiv\frac{\big(M_{\rm T}^{(1)}(x,y)\big)^2}{2p_1p_2}=
\sqrt{\Big[\frac{c_1-x}{r}+y\cos\theta\Big]^2+y^2\sin^2\theta}-
\Big[\frac{c_1-x}{r}+y\cos\theta\Big],
\eeq
\beq\label{eq:z2}
z_2(x,y)\equiv\frac{\big(M_{\rm T}^{(2)}(x,y)\big)^2}{2p_1p_2}=\sqrt{\big[x\cos\theta+(c_2-y)r\big]^2+x^2\sin^2\theta}
-\big[x\cos\theta+(c_2-y)r\big],
\eeq
and $r\equiv p_2/p_1$. It is clear that
\beq
z_1(x,y)\geq0,\quad{\rm and}\quad z_1(x,y)=0\ \Leftrightarrow\ y=0\ \&\ x\leq c_1,
\eeq
\beq
z_2(x,y)\geq0,\quad{\rm and}\quad z_2(x,y)=0\ \Leftrightarrow\ x=0\ \&\ y\leq c_2.
\eeq
From this, we can infer that
\beq
\big(M_{\rm T2}\big)^2=0\quad{\rm if}\quad c_1\geq0\ \&\ c_2\geq0,
\eeq
and for other values of $c_1$ and $c_2$, $\big(M_{\rm T2}\big)^2$ is given by
$\big(M_{\rm T}^{(1)}(x,y)\big)^2=\big(M_{\rm T}^{(2)}(x,y)\big)^2$ at the point $(x,y)=(x_0,y_0)$ 
where the contours of $z_1(x,y)$ and $z_2(x,y)$ in the $x$--$y$ plane become tangent to each other.
We denote the corresponding value $z\equiv z_1(x_0,y_0)=z_2(x_0,y_0)$ in the following.
Eqs. (\ref{eq:z1}) and (\ref{eq:z2}) yield
\beq
x_0=-\frac{r\sin^2\theta}{2z}\bigg(y_0-\frac{z\cos\theta}{\sin^2\theta}\bigg)^2+\frac{rz}{2\sin^2\theta}+c_1,
\eeq
\beq
y_0=-\frac{\sin^2\theta}{2rz}\bigg(x_0-\frac{z\cos\theta}{\sin^2\theta}\bigg)^2+\frac{z}{2r\sin^2\theta}+c_2
\eeq
with the tangential condition
\beq
\frac{\sin^4\theta}{z^2}\bigg(x_0-\frac{z\cos\theta}{\sin^2\theta}\bigg)
\bigg(y_0-\frac{z\cos\theta}{\sin^2\theta}\bigg)=1.
\eeq
We can obtain $z$ by solving these three equations.

A straightforward calculation yields that these equations reduce to
\beq\label{eq:ab}
4(a-b)^2-\big(\sqrt{2(a+b)+3}-1\big)\big(\sqrt{2(a+b)+3}+3\big)^3=0,
\eeq
where
\beq
a=\frac{1}{r^{\frac{1}{3}}}\bigg(\frac{r}{2}-\cos\theta+c_1\sin^2\theta\frac{1}{z}\bigg),\label{eq:app_a}
\eeq
\beq
b=r^{\frac{1}{3}}\bigg(\frac{1}{2r}-\cos\theta+c_2\sin^2\theta\frac{1}{z}\bigg).\label{eq:app_b}
\eeq
It can be checked that the above equations have a unique real positive solution of $z$.
Eqs.~(\ref{eq:ab})-(\ref{eq:app_b}) have been used for the analysis in this work.

In the special case of a ``back-to-back'' pair production, in which Eq. (\ref{eq:BtoB}) holds,
we recover the result obtained by taking the massless limit of the formula in Ref.~\cite{MT2_back2back},
\begin{eqnarray}
\big(M_{\rm T2}\big)^2\Big|_{\rm back-to-back}
=2\big(|\pT{B,1}||\pT{B,2}|+\pT{B,1}\cdot \pT{B,2}\big)
=2p_1p_2(1+\cos\theta).
\end{eqnarray}

\paragraph{(ii) $m_X\ne 0$ case:}
Generalization of the above result for the case with massive $X$, i.e., $m_X\ne0$, is straightforward.
In this case, the $M_{\rm T2}$ variable is defined by Eq.~(\ref{eq:mt2}) with $m_{B}=0$.
The same argument as above shows that Eq.~(\ref{eq:mT2_bound}) holds also in this case.

Calculating in the same way as above, it can be shown that
\beq
\big(M_{\rm T2}\big)^2=m_X^2+2p_1p_2z
\eeq
with $z$ being the solution of Eq.~(\ref{eq:ab}) with
\beq
a=\frac{1}{r^{\frac{1}{3}}}\bigg(\frac{r}{2}-\cos\theta+c_1\sin^2\theta\frac{1}{z}-\frac{r\sin^2\theta}{2}
\frac{m_X^2}{p_2^2}\frac{1}{z^2}\bigg),
\eeq
\beq
b=r^{\frac{1}{3}}\bigg(\frac{1}{2r}-\cos\theta+c_2\sin^2\theta\frac{1}{z}
-\frac{\sin^2\theta}{2r}\frac{m_X^2}{p_1^2}\frac{1}{z^2}\bigg).
\eeq
For the case with massive $X$, this expression for $M_{\rm T2}$ is valid for 
any values of $c_1$ and $c_2$. The existence of a unique positive real solution of $z$ can also be checked.


\begin{thebibliography}{99}

\bibitem{grav10eV}
  H.~Pagels and J.~R.~Primack,
  Phys.\ Rev.\ Lett.\  {\bf 48} (1982) 223;
\\
  M.~Viel, J.~Lesgourgues, M.~G.~Haehnelt, S.~Matarrese and A.~Riotto,
  Phys.\ Rev.\  D {\bf 71} (2005) 063534
  [arXiv:astro-ph/0501562].



\bibitem{Lester-Summers}
  C.~G.~Lester and D.~J.~Summers,
  Phys.\ Lett.\ B {\bf 463} (1999) 99
  [arXiv:hep-ph/9906349];
  \\
  A.~Barr, C.~Lester and P.~Stephens,
  J.\ Phys.\ G {\bf 29} (2003) 2343
  [arXiv:hep-ph/0304226].



\bibitem{Kawagoe:2003jv}
  K.~Kawagoe, T.~Kobayashi, M.~M.~Nojiri and A.~Ochi,
  Phys.\ Rev.\  D {\bf 69} (2004) 035003
  [arXiv:hep-ph/0309031].



  \bibitem{recentMT2}
  C.~G.~Lester and A.~J.~Barr,
  JHEP {\bf0712} (2007) 102
  [arXiv:0708.1028 [hep-ph]];
  \\
  W.~S.~Cho, K.~Choi, Y.~G.~Kim and C.~B.~Park,
  arXiv:0709.0288 [hep-ph];
  \\
  A.~J.~Barr, B.~Gripaios and C.~G.~Lester,
  JHEP {\bf 0802} (2008) 014
  [arXiv:0711.4008 [hep-ph]];
  \\
  W.~S.~Cho, K.~Choi, Y.~G.~Kim and C.~B.~Park,
  JHEP {\bf 0802} (2008) 035
  [arXiv:0711.4526 [hep-ph]].
\\
  M.~M.~Nojiri, Y.~Shimizu, S.~Okada and K.~Kawagoe,
  arXiv:0802.2412 [hep-ph].



\bibitem{ISAJET}
  F.~E.~Paige, S.~D.~Protopopescu, H.~Baer and X.~Tata,
  [arXiv:hep-ph/0312045].
  
  




\bibitem{HERWIG6510}
  G.~Marchesini, B.~R.~Webber, G.~Abbiendi, I.~G.~Knowles, M.~H.~Seymour and L.~Stanco,
  Comput.\ Phys.\ Commun.\  {\bf 67} (1992) 465;
\\
  G.~Corcella {\it et al.},
  JHEP {\bf 0101} (2001) 010
  [arXiv:hep-ph/0011363];
  \\
  G.~Corcella {\it et al.},
  [arXiv:hep-ph/0210213].




  \bibitem{RichterWas:2002ch}
  E.~Richter-Was,
  [arXiv:hep-ph/0207355].



\bibitem{Hamaguchi:2008yu}
  K.~Hamaguchi, E.~Nakamura, S.~Shirai and T.~T.~Yanagida,
  arXiv:0804.3296 [hep-ph].

\bibitem{Allanach:2002nj}
  B.~C.~Allanach {\it et al.},
in {\it Proc. of the APS/DPF/DPB Summer Study on the Future of Particle Physics (Snowmass 2001) } ed. N.~Graf,
{\it In the Proceedings of APS / DPF / DPB Summer Study on the Future of Particle Physics (Snowmass 2001), Snowmass, Colorado, 30 Jun - 21 Jul
2001, pp P125}
  [arXiv:hep-ph/0202233].


  \bibitem{MT2_back2back}
  C.~G.~Lester and A.~J.~Barr, in Ref.~\cite{recentMT2};\\
  W.~S.~Cho, K.~Choi, Y.~G.~Kim and C.~B.~Park, 
  JHEP {\bf 0802} (2008) 035,
  in Ref.~\cite{recentMT2}.

  
\end{thebibliography}
\end{document}